\documentclass{iopart}
\usepackage{iopams,graphics}
\begin{document}

\title[Gauge and constraint degrees of freedom]{Gauge and constraint degrees of freedom:
from analytical to numerical approximations in General Relativity}
%
\author{C. Bona and Dana Alic}\address{Departament de Fisica,
Universitat de les Illes Balears. \\
Institute for Applied Computation with Community Code (IAC3).}
\begin{abstract}
The harmonic formulation of Einstein's field equations is
considered, where the gauge conditions are introduced as dynamical
constraints. The difference between the fully constrained approach
(used in analytical approximations) and the free evolution one
(used in most numerical approximations) is pointed out. As a
generalization, quasi-stationary gauge conditions are also
discussed, including numerical experiments with the gauge-waves
testbed. The complementary 3+1 approach is also considered, where
constraints are related instead with energy and momentum first
integrals and the gauge must be provided separately. The
relationship between the two formalisms is discussed in a more
general framework (Z4 formalism). Different strategies in black
hole simulations follow when introducing singularity avoidance as
a requirement. More flexible quasi-stationary gauge conditions are
proposed in this context, which can be seen as generalizations of
the current 'freezing shift' prescriptions.
\end{abstract}
%

\section{Harmonic formulation}
Einstein's field equations
\begin{equation}\label{Einstein}
  R_{a b} =  8\; \pi\; (T_{a b} - \frac{1}{2}\;T\; g_{a b}),
\end{equation}
are usually expressed as a system of partial differential
equations for the space-time metric $g_{a b}$. Soon after
Einstein's 1915 paper, their mathematical structure was closely
investigated, leading to a very convenient formulation (De Donder
1921, 1927), namely
\begin{equation}\label{DeDonder}
\fl  \frac{1}{2}~g^{cd} \partial^2_{cd} ~g_{ab}
 + \partial_{(a} H_{b)} = \Gamma_{cab}H^c
 + 2~g^{cd}g^{ef}[\partial_e g_{ac}~ \partial_f g_{bd}
 - \Gamma_{ace}~\Gamma_{bdf}] \nonumber \\
 - 8~\pi~(T_{ab} - \frac{T}{2}~g_{ab})\,,
\end{equation}
where indices inside round brackets are symmetrized and we have
noted
\begin{equation}\label{H}
    H^a \equiv - g^{bc}~\Gamma^a_{~bc}
    = 1/\sqrt{g}~\partial_b (\sqrt{g}~g^{ab}).
\end{equation}

One can now take advantage of the general covariance of the
theory. Let us define spacetime coordinates by a set of four
independent harmonic functions, namely
\begin{equation}\label{harmonic}
    \Box~ x^a = 1/\sqrt{g}~\partial_b (\sqrt{g}~g^{ab})=0\,,
\end{equation}
where the box stands for the general-covariant wave operator
acting on functions. In this harmonic coordinate system, the field
equations (\ref{DeDonder}) get the simpler form
\begin{equation}\label{relaxed}
    \Box ~g_{ab}= ~\cdots~ - 16~\pi~(T_{ab} -
    \frac{T}{2}~g_{ab})\,,
\end{equation}
where the dots stand for terms quadratic in the metric first
derivatives. If we look at the principal part (the second
derivatives terms), we see just a set of independent wave
equations, one for every metric component. The coupling comes only
through the right-hand-side terms.

System (\ref{relaxed}) is very convenient in analytical
approximations. Let us assume for instance that the metric admits
a development of the form
\begin{equation}\label{develop}
    g_{ab}=\eta_{ab} + h^{(1)}_{ab} + h^{(2)}_{ab} + \cdots\,,
\end{equation}
where $h^{(n)}$ is the nth-order perturbation. Then, one can
express (\ref{relaxed}) in a recursive way:
\begin{equation}\label{recursion}
\eta^{cd} \partial^2_{cd} ~h^{(n+1)}_{ab} =
F_{ab}(h^{(r)},\partial h^{(s)}) ~~~~~~ r,s\leq n \,,
 \end{equation}
which can by integrated just by inverting the standard
(flat-space) wave operator.

System (\ref{relaxed}), however, is not equivalent to the original
field equations (\ref{DeDonder}). The coordinate conditions
(\ref{harmonic}) can be interpreted as first-order constraints to
be imposed on the solutions of the 'relaxed' system
(\ref{relaxed}). The hard point in proving the well-posedness of
the Cauchy problem for Einstein's equations was precisely to prove
that the harmonic constraints (\ref{harmonic}) where actually
first integrals of the relaxed system (Choquet-Bruhat 1952). In
the analytical perturbation framework, this translates into the
fact that fulfilling the harmonic constraints at the nth-level
does not imply the same thing at the next level. Obtaining a true
solution of the Einstein equations implies adjusting the
integration constants in such a way that
\begin{equation}\label{Hn+1}
    H^{(n+1)}_a \equiv 0 \,,
\end{equation}
and this must be done at every order in the perturbation
development.

\subsection{Numerical Relativity applications}

The relaxed system (\ref{relaxed}) is also very useful in
numerical approximations. The usual practice is using explicit
time-discretization algorithms. This means that the metric
coefficients are computed at a given time slice, assuming that one
knows their values at the previous ones. But again fulfilling the
harmonic constraints (\ref{harmonic}) is not granted. Moreover, in
numerical approximations one has no adjustable integration
constants. This means that numerical errors make the contracted
Christoffel symbols obtained from the relaxed system to depart
from their assumed harmonic (zero) value:
\begin{equation}\label{Ht+dt}
    ^{(relaxed)}\Gamma^a \neq 0\,.
\end{equation}

In this 'free evolution' approach, one can use the non-zero values
(\ref{Ht+dt}) to monitor the quality of the simulation. This can
be done by introducing a 'zero' four-vector $Z^a$ as the
difference between the relaxed and the harmonic (zero) contracted
Christoffel symbols, namely
\begin{equation}\label{Z}
    ^{(relaxed)}\Gamma^a - ^{(harmonic)}\Gamma^a = -2Z^a\,,
\end{equation}
so that true Einstein's solutions would correspond to $Z^a=0$,
which amounts to fulfilling the harmonic constraints. On the
contrary, allowing for (\ref{Z}), solutions of the relaxed system
would verify a generalized version of (\ref{DeDonder}), in which
(\ref{H}) must be replaced by
\begin{equation}\label{HZ}
    H^a \equiv - g^{bc}~\Gamma^a_{~bc}-2Z^a\,.
\end{equation}

The vector $Z^a$ provides a new degree of freedom which arises
quite naturally in numerical simulations. A fully covariant
description is provided by the 'Z4 system' (Bona {\em et al.\/}
2003)
\begin{equation}\label{Z4}
    R_{ab} + Z_{\,a;\,b} + Z_{\,b;\,a} =
    8\pi~(T_{ab}-T/2~g_{ab})~,
\end{equation}
where the semicolon stands for the covariant derivative. We can
easily see that the true Einstein's solutions are recovered when
$Z^a=0$. Moreover, allowing for the contracted Bianchi identities,
the stress-energy tensor conservation implies
\begin{equation}\label{Z4subs}
    g^{bc} Z^a_{~;bc} + R^a_{~b}\,Z^b = 0~.
\end{equation}
It is clear that the vanishing of $Z^a$, leading to true
Einstein's solutions, is a first integral of the 'subsidiary
system' (\ref{Z4subs}), which follows from the field equations
just assuming the stress-energy tensor conservation.

\subsection{Testing coordinate conditions}

We have seen how harmonic coordinates can be preferred for the
sake of simplifying the (approximate) integration of the field
equations. But one would like to use instead the General
Relativity coordinate freedom for simplifying the physical
interpretation of the results. On the contrary, choosing a
convoluted coordinate system can complicate even the simplest
physical situations.

A good example of these unphysical gauge complications is the
'gauge waves' testbed (Alcubierre \etal 2004). The Minkowsy (flat)
metric can be written in some non-trivial harmonic coordinate
system as:
\begin{equation}\label{gaugew}
    ds^2 = F(x-t)(-dt^2 + dx^2) + dy^2 + dz^2 ~,
\end{equation}
where $F$ is an arbitrary function of its argument. One could
naively interpret this as the propagation of an arbitrary wave
profile with unit speed. But it is a pure gauge effect, because
(\ref{gaugew}) is nothing but the Minkowsky metric in disguise. A
more natural coordinate system should be adapted to the fact that
flat spacetime is stationary, and this is not granted by the
harmonic condition, as (\ref{gaugew}) shows dramatically.

The problem of finding 'quasi-stationary coordinates' (as
stationary as possible) in a generic spacetime has been addressed
recently (Bona \etal 2005a). The idea is to find 'almost-Killing'
vector fields $\xi^a$ by means of a standard variational principle
\begin{equation}\label{action}
    \delta S = 0~, ~~~~  S \equiv \int L ~\sqrt{g} ~d^4 x~\,,
\end{equation}
where the Lagrangian density $L$ is given by
\begin{equation}\label{Lag1}
    L = \xi_{(a\,;\,b)}\xi^{(a\,;\,b)} -
    \frac{k}{2}~(\xi^c_{~;\,c})^2~
\end{equation}
($k$ being an arbitrary parameter), and the variations of the
vector field $\xi$ are considered in a fixed spacetime. The
resulting Euler-Lagrange equations get the form
\begin{equation}\label{AKE}
     [~\xi^{\,a;\,b} + \xi^{\,b;\,a}
     - k~\xi^c_{~;\,c}~g^{ab}~]_{;\,b}=0~.
\end{equation}
('almost-Killing' equation), or the equivalent one
\begin{equation}\label{AKE2}
    g^{bc}\xi_{a;bc} + R_{ab}\,\xi^b
    + (1-k)~\partial_a (\xi^c_{~;\,c})= 0~.
\end{equation}

We will consider here the particular 'harmonic' choice $k=1$. Note
that, in this case, the subsidiary system (\ref{Z4subs}) is
nothing but condition (\ref{AKE2}) for the $Z^a$ vector. We can
then interpret that the combination $Z_{(a\,;\,b)}$ in the Z4
system (\ref{Z4}) gets minimized, so that one gets as close as
possible to the original Einstein system. The name 'harmonic' for
the $k=1$ choice can be justified if we take the integral curves
of $\xi$ to be the time lines of our coordinate system. In this
'adapted coordinate system', condition (\ref{AKE}) reads simply
\begin{equation}\label{Gamma}
    g^{bc} \partial_t \Gamma^a_{~bc} = 0~\,,
\end{equation}
which is a close generalization of the harmonic coordinates
condition (see Bona \etal 2005a for more details).

In order to test the behavior of these quasi-stationary
conditions, we will consider the 'gauge waves' form (\ref{gaugew})
of the flat metric, with the following profile:
\begin{equation}\label{sinusw}
    F = 1 - A~Sin(~2\pi (x-t)~)~~~~~~A=0.1\,,
\end{equation}
so that the resulting metric is periodic and we can identify for
instance the points $-0.5$ and $0.5$ on the $x$ axis. This allows
to set up periodic boundary conditions in numerical simulations,
so that the initial profile keeps turning around along the $x$
direction. Note that the coordinate conditions (\ref{AKE}) require
the use of some damping term in order to ensure the stability of
the solutions (see Bona \etal 2005b for details).

\begin{figure}
\centering
\scalebox{0.5}[0.6]{\includegraphics{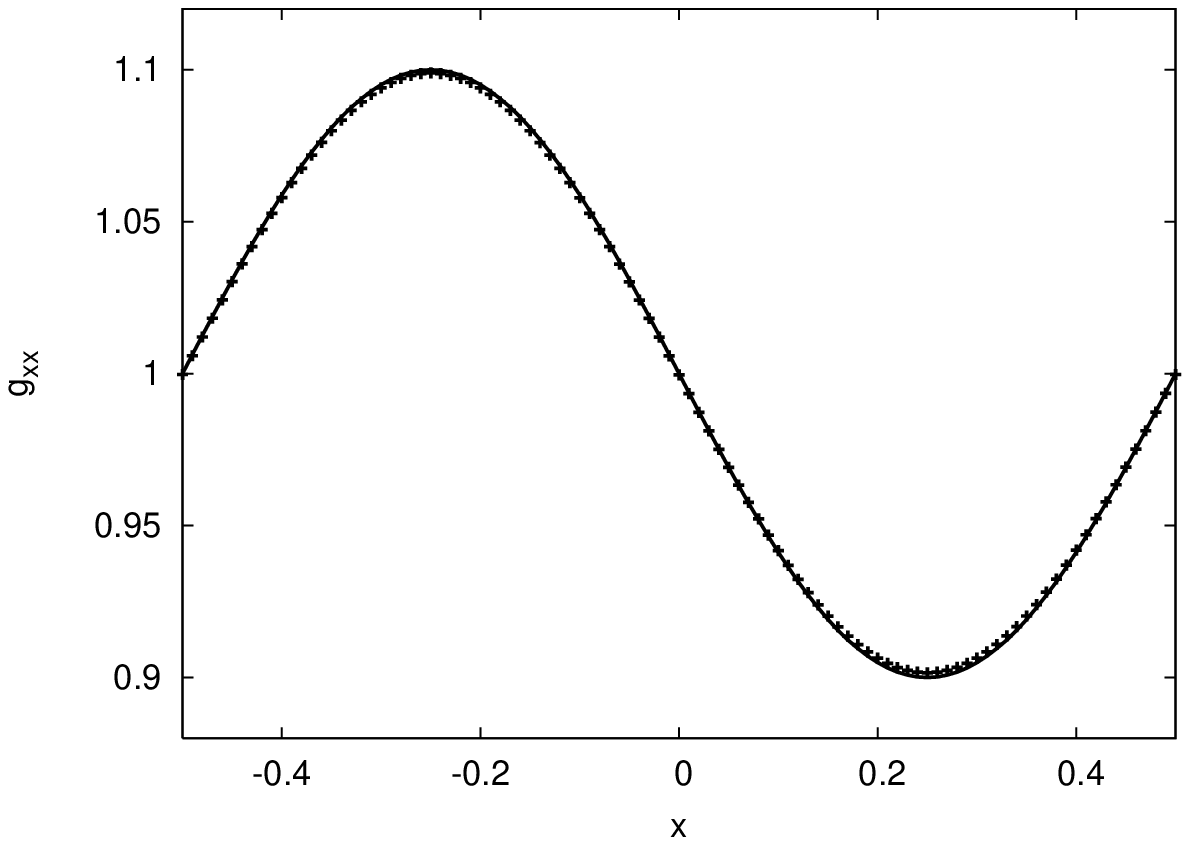}
\includegraphics{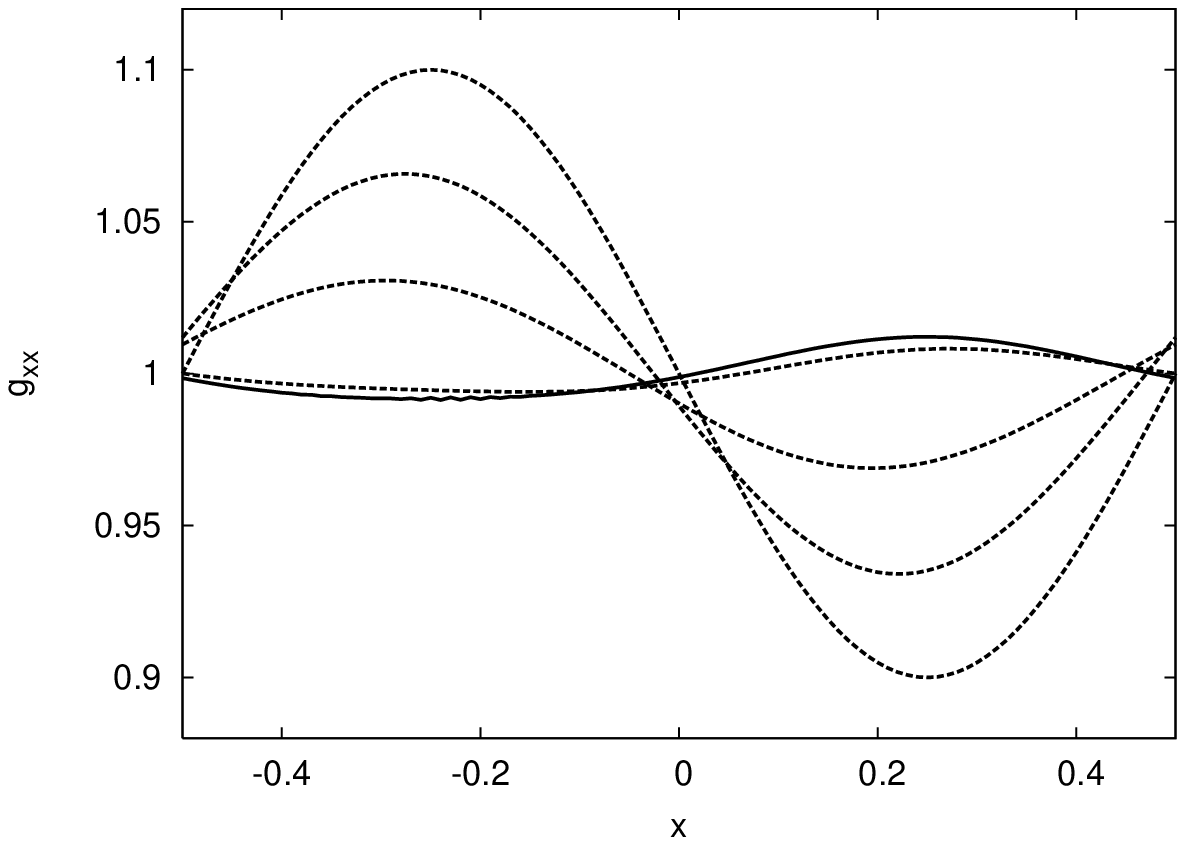}}
\caption{Gauge waves simulation with periodic boundary conditions
and sinusoidal initial data for the metric component $g_{xx}$. The
left panel corresponds to the harmonic coordinates case. After
$100$ round trips, the evolved profile (cross marks) nearly
overlaps the initial one (continuous line), which corresponds also
with the exact solution in the harmonic case. The right panel
corresponds to the same simulation for the quasi-stationary
coordinates defined by (\ref{Gamma}). We show the initial data and
the evolved profiles after $1$, $2$, $5$ and $100$ round trips
(continuous line). The solution clearly approaches the Minkowsky
value ($g_{xx}=1$), although a small residual profile remains.}
\label{gaugewplot}
\end{figure}

The results of the numerical simulations are displayed in
Fig.~\ref{gaugewplot}. The left panel shows the harmonic case,
where the numerical results follow very closely the exact
solution. Only a small amount of numerical dissipation is visible
after $100$ round trips (we are using here a third-order-accurate
finite volume method in order to get rid of the dominant
dispersion error). The right panel shows the behavior of the
quasi-stationary condition (\ref{Gamma}) for the same initial
data. We see that the amplitude is quickly decreasing, so that we
get very close to the stationary (Minkowsky) value $g_{xx}=1$
after only $5$ round trips. Although a small residual profile
remains, even after $100$ round trips (continuous line), the
initial amplitude has been reduced by an order of magnitude.

\section{3+1 formulation}

About thirty years after the advent of the harmonic formalism, the
'evolution formalism' (Lichnerowicz 1944, Choquet 1956 provided a
breakthrough in the understanding of Einstein's equations
structure. Four-dimensional spacetime was described as sliced by
constant-time hypersurfaces. Space-time dynamics was then
described as the time evolution of the three-dimensional geometry
of these surfaces. This '3+1 formalism' was very convenient at
that time for studying the initial value problem and it is widely
used today in numerical applications.

The four-dimensional metric can be adapted to the 3+1 geometry in
the following way:
\begin{equation}\label{line3+1}
    ds^2 = g_{ab}~dx^adx^b = - \alpha^2\;dt^2
    + \gamma_{ij}\;(dx^i+\beta^i\;dt)\;(dx^j+\beta^j\;dt),
\end{equation}
where $\gamma_{ij}$ is the three-dimensional metric on every
slice. The 'lapse function' $\alpha$ measures the proper-time
versus coordinate-time ratio when moving along the lines normal to
the slices ($\beta^i=0$). We will see how to take advantage of
this degree of freedom in what follows. The 'shift' $\beta^i$
measures in turn the deviation between these normal lines and the
time lines. It does not affect the geometry of the slicing in any
way. The extrinsic curvature of the slices is the proper time
derivative of the space metric along the normal lines, namely:
\begin{equation}\label{K}
  K_{ij} = -1/2\alpha~(\partial_t -{\cal L}_{\beta})~
  \gamma_{ij}\,,
\end{equation}
where $\cal L$ stands for the Lie derivative.

In this framework, the set of ten Einstein's field equations
(\ref{Einstein}) can be decomposed into two different subsets (see
Choquet 1967 for details). The space components provide six
first-order evolution equations for $K_{ij}$ (amounting to
second-order evolution equations for $\gamma_{ij}$). The
components
\begin{equation}\label{constraints}
    n_b~(G^{ab}= 8~ \pi~ T^{a b})
\end{equation}
(where $G_{ab}\equiv R_{ab}-\frac{1}{2}\;R\; g_{a b}$ is the
Einstein tensor and $n_a$ is the field of unit normals to the time
slicing) provide instead four constraint equations for the pair
$\gamma_{ij}$, $K_{ij}$. These 'energy-momentum constraints' arise
just from the field equations, independently of the coordinate
gauge. Moreover, one does not get here any evolution equation for
the lapse or the shift, which are just kinematical quantities. In
order to determine them, one must specify four coordinate
conditions, which must be added to the field equations in order to
complete the evolution system. One can provide for instance four
extra evolution equations
\begin{equation}\label{evolQ}
    (\partial_t-\beta^k\partial_k) \alpha = -\alpha^2 Q ~,~~~~
    (\partial_t-\beta^k\partial_k)\beta^i = -\alpha Q^i~,
\end{equation}
where $Q$ and $Q^i$ can be freely specified.

In order to see the relationship between the harmonic and the 3+1
formalisms, let us decompose the contracted Christoffel symbols
$\Gamma^a \equiv g^{bc}\Gamma^a_{bc}$, namely:
\begin{equation}\label{GammaQs}
    n_a\Gamma^a = \alpha~\Gamma^0 = Q-trK ~,~~~~
    \alpha~\Gamma_i = Q_i - \partial_i\alpha + \alpha~
    ^{(3)}\Gamma_i~.
\end{equation}
It follows that fixing the value of $\Gamma^0$ amounts to provide
the evolution equation for the lapse (which determines the time
slicing), whereas fixing the value of $\Gamma_i$ amounts to
provide the evolution equation for the shift (which determines the
time lines for a given slicing). The main difference is that the
coordinate conditions (\ref{harmonic}) where introduced as
constraints in the harmonic formalism, whereas the corresponding
3+1 conditions (\ref{evolQ}) are part of the evolution system.
This will have important consequences in what follows.

\subsection{Gravitational collapse scenarios}

Gravitational collapse in General Relativity can lead to the
arising of spacetime singularities, even from regular initial data
(a massive star for instance). This is a serious issue in
Numerical Relativity, as the computation can not proceed beyond
singularity formation, even if the outside spacetime regions are
regular. The lapse degree of freedom can be used in these cases to
allow the computation to continue (in coordinate time) by locally
diminishing the lapse (and then the proper time flow) in the
collapsing regions, where the space volume element
$\sqrt{\gamma}\,$ is diminishing. When done properly, the time
slices do not reach the collapse singularity in a finite amount of
coordinate time (singularity avoidance).

Let us illustrate this behavior by considering the harmonic
slicing condition. Allowing for (\ref{GammaQs}), this means to
take $Q=trK$. As far as singularity avoidance is a geometrical
property of the slicing, the value of the shift is irrelevant, so
we will use normal coordinates (zero shift) for simplicity. This
condition can be easily integrated: allowing for the definitions
(\ref{K}, \ref{evolQ}) we get
\begin{equation}\label{harmslicing}
    \partial_t~(\sqrt{\gamma}/\alpha) = 0 ~~~~
    \Rightarrow ~~~~ \alpha \sim \sqrt{\gamma}~.
\end{equation}
It follows that the lapse tends to vanish locally (collapse of the
lapse) where the space volume element goes to zero. Although one
gets in this way arbitrarily close to the singularity, it can be
shown that one does not actually reach it in a finite amount of
coordinate time (Bona and Mass\'{o} 1988).

The singularity avoidance of the harmonic slicing is really a
limit case, vulnerable to numerical errors. This is a problem in
gravitational collapse scenarios, specially in the harmonic
formalism, where we must remember that the harmonic conditions
where not part of the relaxed system: they were just dynamical
constraints. This is why numerical codes based in the harmonic
formalism currently excise the singularity-forming regions out of
the computational domain. This amounts to set up an artificial
internal boundary in the strong field region. Moreover, this must
be a dynamical boundary, capable of changing shape or moving
across the numerical grid. This is a difficult issue, but still
feasible. This has been achieved for instance in the most recent
Numerical Relativity breakthrough, where a binary black hole
simulation lasted long enough for extracting a consistent
gravitational wave signal for the first time (Pretorius 2005).

An obvious alternative to dynamical excision is to consider time
slicing conditions with stronger singularity-avoidance properties.
This can be done by generalizing the harmonic slicing condition in
the following way (Bona \etal 1995):
\begin{equation}\label{genharmonic}
    Q = f~tr\,K~,
\end{equation}
where $f$ is an arbitrary function of the lapse, so that the
original harmonic slicing is recovered for $f=1$. Most of the
current binary-black-hole simulations in the BSSN formalism
actually use the slicing condition (\ref{genharmonic}) with the
choice $f=2/\alpha$. This corresponds to the '1+log' slicing
condition (Bernstein 1993), the name coming from the resulting
form of the lapse in normal coordinates (zero shift):
\begin{equation}\label{1+log}
    \alpha = \alpha_0 + ln(\gamma/\gamma_0)~.
\end{equation}
It follows from (\ref{1+log}) that the coordinate time evolution
stops before even getting close to the collapse singularity. It is
easy to see in this case that the time evolution has a fixed point
where the lapse vanishes (a limit surface). In normal coordinates,
this happens when
\begin{equation}\label{limitsurf}
    \gamma = \gamma_0~exp\,(-\alpha_0)~,
\end{equation}
that is well before the vanishing of the space volume element (the
initial lapse value being usually close to one). The appearance of
a limit surface provides a safety margin for black hole
simulations, as far as numerical errors have no chance to result
into hitting the singularity.

Unfortunately, singularity-avoidance does not come for free. Time
slices get distorted in the process, mainly by a length increase
along the radial direction (slice stretching, see Reimann and
Br\"{u}gmann 2004), which is nevertheless compatible with the
overall collapse. This causes a progressive loss of resolution
that can produce high-frequency noise in numerical simulations
where the grid gets too coarse to resolve the profiles of the
dynamical fields. This is analogous to the well known Gibbs
phenomenon, and usually leads to code crashing as far as the noise
starts growing quickly.

Slice stretching is inherent to singularity avoidant slicing
conditions, but the appearance of high-frequency noise can be
delayed, even avoided, in many ways. One can just increase the
grid resolution or use instead more specific tools: artificial
dissipation terms (Gustavson \etal 1995) or advanced finite volume
methods, like the ones currently used in Computational Fluid
Dynamics (for a recent implementation, see Alic \etal 2007). Just
after Pretorius paper, many groups published improved
binary-black-hole simulations, obtained with 3+1 codes using the
'1+log' slicing condition (Campanelli \etal 2006, Baker \etal
2006, Diener \etal 2006).

\subsection{Quasi-stationary gauge conditions and singularity avoidance}

At this point, it is interesting to ask whether the
quasi-stationary coordinate conditions derived  from the
variational principle (\ref{action}) can be made compatible with
the singularity avoidance requirement. Note that, when adapting
our spacetime coordinates to a solution $\xi$ of the
almost-Killing equation (\ref{AKE}), we are demanding two
different things: the time lines are chosen to be the integral
curves of $~\xi~$ and the time coordinate is chosen to be the
preferred affine parameter associated to these lines. Although the
first requirement fits perfectly into the idea of getting a
quasi-stationary gauge condition, the second one lacks of a clear
physical motivation. It seems that singularity avoidance is not
enforced in this way.

A better strategy is to choose a priori the time coordinate, that
is a spacetime slicing given by
\begin{equation}\label{slicing}
    \phi(x^a) = {\it constant}~,
\end{equation}
with a view to enforce singularity avoidance. Then, we can use
this time coordinate as a parameter along the integral lines of
the almost-Killing vector $\,\xi\,$, by requiring
\begin{equation}\label{xinorm}
    \xi^a~\partial_a~\phi = 1~.
\end{equation}
This amounts to constraint the vector $\,\xi\,$ to fulfill
(\ref{xinorm}) in the minimization process. In other words, we
introduce a Lagrange multiplier and minimize
\begin{equation}\label{Lagprime}
    L' \equiv L + \lambda~(\xi^a\partial_a~\phi - 1)
\end{equation}
(for any $~\phi~$ given a priori) instead of the original
Lagrangian (\ref{Lag1}).

The resulting Euler-Lagrange equations include now the constraint
(\ref{xinorm}), plus the system
\begin{equation}\label{ELeqs}
     [~\xi^{\,a;\,b} + \xi^{\,b;\,a}
     - k~\xi^c_{~;\,c}~g^{ab}~]_{;\,b}=\lambda~\partial^a\phi~.
\end{equation}
which generalizes the almost-Killing equation (\ref{AKE}). Using
adapted coordinates,
\begin{equation}\label{xiadapted2}
    \phi = t~, \qquad \xi = \partial_t~,
\end{equation}
it can be written as
\begin{equation}\label{ELadap}
    g^{bc}\partial_t\Gamma^a_{~bc}
    +(1-k)~g^{ab}\partial_t\Gamma^c_{~bc}=\lambda~g^{at}~.
\end{equation}

Let us now split condition (\ref{ELadap}) into its 3+1 components.
The time component can be ignored, because it just provides the
value of the Lagrange multiplier itself. Remember that the time
slicing was chosen a priori, so that we do not need any further
condition for the lapse. The (downstairs) space component,
\begin{equation}\label{shift}
        g_{ia}\,g^{bc}\partial_t\Gamma^a_{~bc}
    +(1-k)~\partial_t\Gamma^c_{~ic}=0~,
\end{equation}
provides a (second order) evolution equation for the shift. In
this way, we have completely splitted the gauge conditions: the
shift equation (\ref{shift}) should work with any time slicing,
that can be freely chosen a priori.

The quasi-stationary shift condition (\ref{shift}) is completely
independent of the value of the Lagrange multiplier. This means
that we would get the same condition from the original
unconstrained Lagrangian. We can conclude that the slicing
constraint (\ref{xinorm}) does  not affect the minimization
process in the shift sector.

Let us test these conditions in a spherically symmetric black-hole
simulation. We will write the Schwarzschild line element in the
'wormhole' form:
\begin{equation}\label{ds2}
    {\rm d}s^2 = -(~tanh\, \eta~)^2~dt^2
    + 4m^2~(~cosh\, \eta/2~)^4~(~d\eta^2 + d\Omega^2~)~,
\end{equation}
which can be obtained from the isotropic form by the following
coordinate transformation
\begin{equation}\label{eta}
    r = m/2~exp\,(~\eta~)~.
\end{equation}

\begin{figure}
\centering \scalebox{0.9}[0.6]{\includegraphics{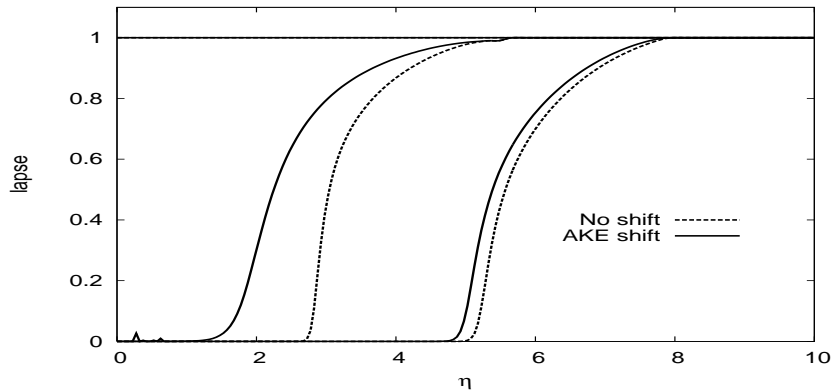}}
\caption{Plots of the lapse function evolution for two spherically
symmetric (1D) black-hole simulations. The 1+log slicing condition
has been used in both cases, producing the lapse collapse. The
space coordinates, however, are different: we compare normal
coordinates (no shift) with quasistationary coordinates (AKE
shift). The shift delays the collapse, as it provides an outgoing
speed for the grid nodes. A smoothing of the slopes can also be
seen as a side effect. The logarithmic character of the grid makes
these effects less apparent for larger values of the $\eta$
coordinate.} \label{lapse}
\end{figure}

We will combine the '1+log' slicing prescriptio with the
quasi-stationary shift condition (\ref{shift}) with $k=1/2$. We
run the simulation up to $1000m$, as in the zero shift case
(normal coordinates). We see in Fig.~\ref{lapse} that the lapse
collapses as usual, avoiding the singularity in both cases as
expected. The effect of the shift is adding some outgoing speed to
the grid nodes, so that the advance of the collapse front across
the grid is delayed. We have added to the shift condition
(\ref{shift}) a standard damping term (Bona \etal 2005b) in order
to avoid the shift values to grow without control. Note that the
slopes at the collapse front are smoothed out, so they can be
better resolved. The logarithmic character of the grid makes these
effects less apparent for larger values of the $\eta$ coordinate.

Our results prove that the quasi-stationary shift conditions
(\ref{shift}) are actually compatible with standard
singularity-avoidant slicing conditions. We have found, however,
that the results can depend crucially of particular value of the
parameter $k$. It is suggestive that, at least in the particular
case shown here, the best results are obtained with the  $k=1/2$
choice. This is a very special value, because the minimum
principle (\ref{action}) leads in this case to a minimisation of
the conformal-Killing equation: a quasi-conformal shift condition.
This opens an interesting perspective for future work.

\section*{Acknowledgements}
{\em This work has been supported by the
Spanish Ministry of Science and Education through the research
project number FPA2004-03666 and by the Balearic Conselleria
d'Economia Hissenda i Innovaci\'{o} through the project
PRDIB-2005GC2-06}.

\References


\item[] Alcubierre, M. {\em et al} 2004,
    {\it Class. Quantum Grav.} 21(2), 589–613.
\item[] Alic, D. Bona, C. Bona-Casas, C. and Mass\'o, J.
    2007, {\it Phys. Rev.} D (in press), arXiv:0706.1189.
\item[] Baker, J.G. {\it et al} 2006,
    {\it Phys. Rev. Lett.} 96, 111102.
\item[] Bernstein, D. 1993, {\it Ph.~D.~Thesis}
    (Dept.~of Physics, Univ.~of Illinois at Urbana-Champaign).
\item[] Bona, C. and Mass\'o, J. 1988,
    {\it Phys. Rev.} D38, 2419.
\item[] Bona, C. Mass\'o, J. Seidel, E. and Stela,
    J. 1995, {\it Phys. Rev. Lett.} 75, 600.
\item[] Bona, C. Ledvinka, T. Palenzuela, C.
    and \v Z\'a\v cek, M. 2003, {\it Physical Review} D67, 104005.
\item[] Bona, C. Carot, J. and Palenzuela-Luque, C.
    2005, {\it Phys. Rev.} D72, 124010.
\item[] Bona, C. Lehner, L. and Palenzuela-Luque, C.
    2005, {\it Phys. Rev.} D72, 104009.
\item[] Campanelli, M. Lousto, C.O.
    Marronetti, P. and Zlochower, Y.  2006,
    {\it Phys. Rev. Lett.} 96, 111101.
\item[] Four\'{e}s-Bruhat, Y. 1952,
    {\it Acta Math.} 88, 141.
\item[] Choquet-Bruhat, Y. 1956,
    {\it J. Rat. Mec. Analysis} 5, 951.
\item[] Choquet-Bruhat, Y. 1967, in
    {\it Gravitation: An Introduction to Current Research (L. Witten,
    editor).} John Wiley, New York.
\item[] De Donder, T. 1921,
    {\it La Gravifique Einstenienne}, Gauthier-Villars, Paris.
\item[] De Donder, T. 1927,
    {\it The Mathematical Theory of Relativity}, Massachusetts
    Institute of Technology, Cambridge, MA.
\item[] Diener, P. {\it et al} 2006,
    {\it Phys. Rev. Lett.} 96, 121101.
\item[] Gustafson, B. Kreiss, H.O. and Oliger, J.
    1995, {\it Time dependent problems and difference methods}
    (New York: Wiley).
\item[] Lichnerowicz, A. 1944,
    {\it J. Math. Pures Appl.} 23, 37.
\item[] Pretorius, F. 2005,
    {\it Phys. Rev. Lett.} 95, 121101.
\item[] Reimann, B. and Br\"{u}gmann, B. 2004,
    {\it Phys. Rev.} D69 044006.

\endrefs
\end{document}